\documentclass[10pt,conference]{IEEEtran}
\usepackage{graphicx} 
\usepackage{cite}
\usepackage{url}

\title{Quantum approaches for particle-in-cell codes}

\author{\IEEEauthorblockN{Ryan J. J. Connor\IEEEauthorrefmark{1}, Andrew J. Daley\IEEEauthorrefmark{1}, Callum W. Duncan\IEEEauthorrefmark{2}, Preetma Soin\IEEEauthorrefmark{3}}
\IEEEauthorblockA{\IEEEauthorrefmark{1} Department of Physics, University of Oxford, Clarendon Laboratory, OX1 3PU Oxford, United Kingdom}
\IEEEauthorblockA{\IEEEauthorrefmark{2} Aegiq Ltd.,Cooper Buildings, Arundel Street, Sheffield S1 2NS, United Kingdom}
\IEEEauthorblockA{\IEEEauthorrefmark{3} AWE, Aldermaston, Reading, RG7 4PR, United Kingdom\\
Email: ryan.connor@physics.ox.ac.uk}}

\makeatletter
\def\ps@IEEEtitlepagestyle{%
\def\@oddfoot{\mycopyrightnotice}%
\def\@evenfoot{\mycopyrightnotice}%
}
\def\mycopyrightnotice{%
{UK Ministry of Defence \copyright Crown owned copyright ~\copyright2026 /AWE \hfill} 
\gdef\mycopyrightnotice{}
}

\begin{document}

\maketitle

\begin{abstract}
    Solving the high-dimensional, nonlinear partial differential equations that govern the rich physics of plasmas remains a major computational challenge. In fluid dynamics, quantum-inspired methods based on tensor-network techniques have recently emerged as a promising new paradigm, offering large degrees of data compression on conventional hardware and a natural pathway towards solving exponentially large problems on quantum computers. Early studies suggest that tensor-network methods may also help overcome key bottlenecks in computational plasma dynamics, particularly for industry-relevant scenarios where conventional simulations remain costly. In this work, we present recent progress in developing tensor-network methods compared to standard particle-in-cell codes and discuss their capability to overcome the bottlenecks of conventional approaches. We further extend the tensor-network formulation to support the insertion of electromagnetic fields at arbitrary spatial locations. This capability is demonstrated through the injection of a laser field to accelerate a plasma, providing a key step towards tensor-network-based simulation of externally driven plasma systems.
\end{abstract}

\section{Introduction}

Tensor-networks are a widely used tool in quantum-many body physics \cite{biamonte2017tensornetworksnutshell,TN_quantum_systems} where they have been used to great success to probe low lying energy states through the seminal density matrix renormalisation group algorithm\cite{DMRG}, and to study out of equilibrium phenomena via performing time-evolution of an initial quantum state under the action of a Hamiltonian  \cite{Paeckel_2019,Vanderstraeten_2019,TDVP,DMRG}.  Tensor-networks provide a mechanism to achieve physically motivated data compression, allowing the simulation of many-interacting quantum particles by limiting the entanglement in the representation. 

Recently, there has been an emerging focus on extending the application of tensor-networks to simulations of non-linear partial differential equations \cite{Q_inspired_fluids,wall_bounded_flows,holscher2024quantuminspiredfluidsimulation2d,multigrid,GPE_paper,gourianov2024tensornetworksenablecalculation}. One particular area of promise for the use of tensor-networks is that of simulations of plasma dynamics \cite{ye2024quantized,Plasma_Connor}. Understanding and predicting the complex phenomena which occur when a plasma interacts with electric and magnetic fields is a central task across many disciplines which remains challenging for large 2D and 3D problems. The de facto description of plasmas is that of the Vlasov-Maxwell kinetic model, where the plasma is described by a distribution function in phase space $f_s(\mathbf{x},\mathbf{v},t)$,

\begin{equation}
    \frac{\partial f_s}{\partial t} + \mathbf{v}\cdot \nabla_r f_s \nonumber + \frac{q_s}{m_s} [\mathbf{E}+\mathbf{v}\times\mathbf{B}]\cdot \nabla_v f_s=0 \ , \label{eq:Vlasov}
\end{equation}
where $\mathbf{E}$ and $\mathbf{B}$ are the electric and magnetic fields which are updated via Maxwells equations.

A standard standard approach for simulations of Eq.~\ref{eq:Vlasov} is based on particle-in-cell (PIC) codes. PIC methods are a hybrid approach involving computational particles and a fixed grid upon which the electric and magnetic fields are specified. These computational particles are “macroparticles” which represent a number of smaller similar physical particles. The basic PIC loop begins first with the particles being “pushed” according to Lorenz equations based on the electric and magnetic field properties. Then particle deposition quantities such as charges and current are calculated. Based on these quantities the electric and magnetic field values are solved on the PIC grid using Maxwell’s equations. The cycle then repeats itself and the particles are pushed with the newly interpolated electric and magnetic field values. 

\begin{figure}[t!]
    \centering
    \includegraphics[width=0.6\linewidth]{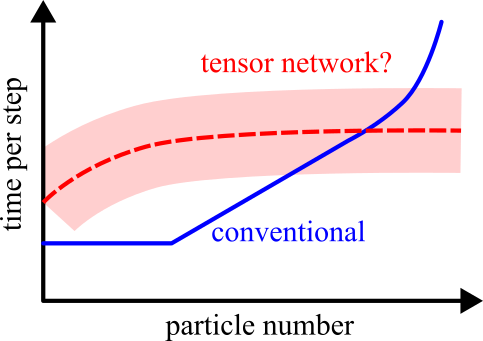}
    \caption{Illustration of time complexity. Conventional PIC approaches (solid blue line) are informed by runs of real problems, and show the emergence of the linear scaling with particle number with initial offset set by the grid resolution. The desired scaling of tensor network approaches (dashed red line) is logarithmic with a larger initial offset. The shaded region is to indicate the uncertainty in the offset, which is problem and implementation dependent.}
    \label{fig:scaling}
\end{figure}

PIC codes are widely used for applications such as laser-plasmas interactions, ion beam propagation through a plasma, and beyond \cite{LI2023111733,ZHANG2024108932,Beat_paper,Nastac_2024}. Modelling these applications in PIC is subject to constraints both on the grid spacing and on the particles. The spatial scale of the grid is subject to being less than or equal to the Debye length otherwise the shielding effects of the electric field will not be correctly captured. Small grids with fixed particle number, as well as statistical noise which scales as the inverse of the particle number, results in artificial self-heating in PIC codes which in turn feeds instabilities. Mitigation strategies for this include increasing the particle number with a finer grid but this increases computational complexity and simulation time. The time complexity of PIC codes scales with both the grid resolution and the particle number, we illustrate the typical scaling in Fig.~\ref{fig:scaling}, where we fix the grid resolution and scale the particle number. This results in a fixed offset until the particle number dominates the scaling and we see the emergence of a linear scaling. For very large particle numbers, memory and system limits dominate and we get the emergence of an exponential scaling.

In contrast to PIC codes, tensor-networks can directly solve Eq.~\ref{eq:Vlasov} by discretisation of the full phase space grid without the need for computational particles, provided the discarded correlations do not impact the underlying physics. One can therefore eliminate the statistical noise and self-heating present in PIC. If designed well, and within the approximation of limited correlations, the time complexity of tensor network approaches scale logarithmically with system size, but with a higher initial computational cost. We illustrate in Fig.~\ref{fig:scaling} the hoped for scaling in comparison to conventional methods, which motivates the ambition of this work to unlock larger scale simulations than is capable with current PIC codes.

Tensor-networks are intimately linked to quantum circuits, including the emulation of quantum hardware \cite{berezutskii2025tensornetworksquantumcomputing,Tindall_2024}. By understanding the regimes in which tensor-network methods succeed and fail, we increase our knowledge of how one may be able to design and implement full quantum algorithms for plasma dynamics as well as enabling study of quantum algorithms implemented at large scales.

\section{Overview of tensor-networks for plasmas}

\begin{figure}[t!]
    \centering
    \includegraphics[width=0.8\linewidth]{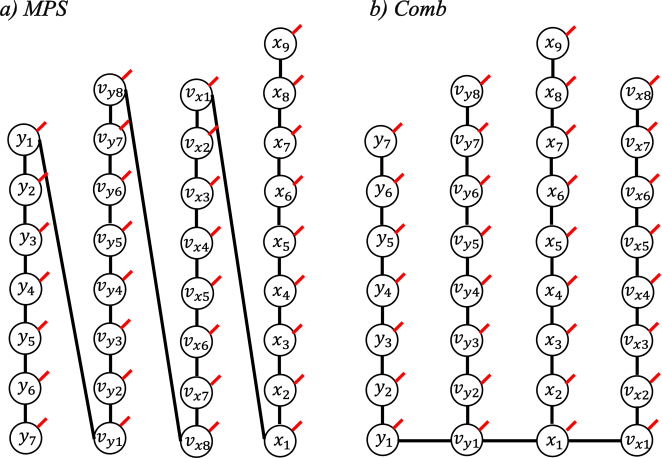}
    \caption{Illustration of two tensor-network encodings of $f(x,y,v_x,v_y)$ on an $2^9\times 2^7\times2^8\times2^8$ phase space grid into an $a)$ MPS and $b)$ comb tensor-network geometry. We choose to label each tensor according to which dimension it belongs to, and the corresponding length scale of that dimension. E.g. $x_1$ encodes the coarsest length scale of the $x$ dimension, and $v_{y_8}$ encodes the finest length scale of the $v_y$ dimension. Bond indices are shown as black lines connecting two tensors, whilst the physical indices are shown in red emerging diagonally upwards from each tensor. }
    \label{fig:Beat_encodings}
\end{figure}

To simulate the Vlasov equation with tensor-networks we first must represent the phase space function $f_s(\mathbf{x},\mathbf{v})$ of Eq.~\ref{eq:Vlasov} as a tensor-network state. To do so, unless otherwise stated, we make use of a sequential encoding into matrix product state (MPS) format, which means we encode spatial dimensions as connected segments of a larger MPS, as illustrated in Fig.~\ref{fig:Beat_encodings}a. To do this, each parameter $l\in\{x,y,v_x,v_y \cdots\}$ is discretised into a grid of $2^{N_l}$ grid points labelled $q_i^l$, with $i\in \{0,1,2,\cdots,2^{N_l}-1\}$. Each grid-point can be written in a binary representation,
\begin{equation}
    q_i^l = \left( \sigma_1^l, \sigma_2^l, \dots, \sigma_{N_l}^l \right),
\end{equation}
where $\sigma_j^l\in \{0,1\}$. Expressing $f_s$ in terms of these binary coefficients and applying singular value decompositions gives the MPS representation of tensors $M^{\sigma_j^l}_{\alpha_n}$,
\begin{equation}
    f_s = \sum_{\alpha_1}^{\chi_1} \sum_{\alpha_2}^{\chi_2} \dots \sum_{\alpha_{N-1}}^{\chi_{N-1}} M_{\alpha_1}^{{\sigma}_1} M_{\alpha_1,\alpha_2}^{{\sigma}_2} \dots M_{\alpha_{N-1}}^{{\sigma}_{N}} \ , \label{eq:MPStensor}
\end{equation}
with $\alpha_n$ the indices of MPS bonds, and $\chi_n$ the $n$th bond dimension of the MPS. The $\chi_n$ encode correlations between tensors. The vectors $\mathbf{v}, \mathbf{E}$ and $\mathbf{B}$ are encoded as MPSs in the same manner. The differential operators $\nabla_r$ and $\nabla_v$ can be encoded exactly as tensor-network operators \cite{Plasma_Connor}.

\begin{figure*}[t!]
    \centering
    \includegraphics[width=\linewidth]{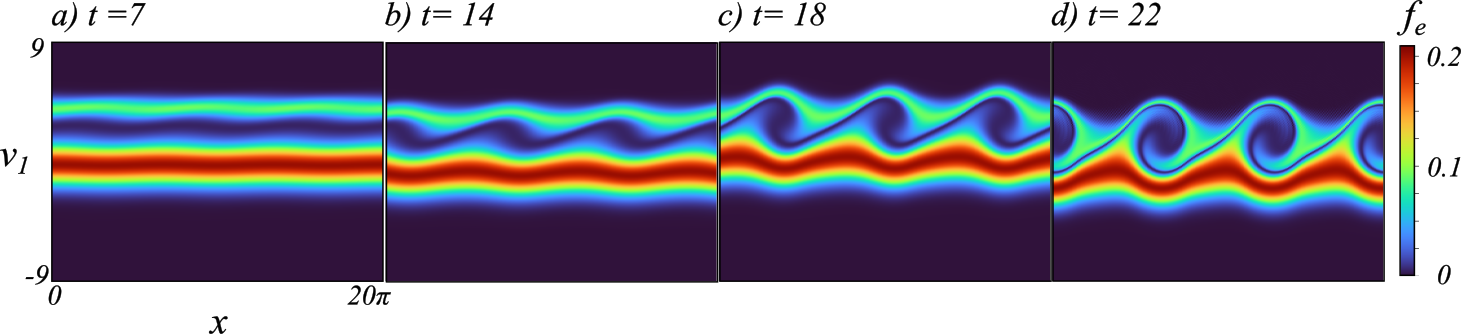}
    \caption{Snapshots of a cut of the electron distribution $f_e(x,v_1)$ through phase space for the bump-on-tail instability reproduced from \cite{Plasma_Connor}. We show a slice through the $v_2=0$ plane, resulting in a 2D function in $x$ and $v_1$, at various simulation times of  $a) t=7$, $b) t=14$, $c) t=18$ and $d) t=22$. These simulations where conducted on a $2^8 \times 2^8 \times 2^8$ phase space grid, resulting in a $24$ site MPS encoding, and evolved via a fixed time-step $4th$ order Runge-Kutta scheme with a time-step of $dt=0.0014$. A truncation cutoff of $10^{-15}$ was used to adaptively control the bond dimension $\chi$. }
    \label{fig:Bump_snaps}
\end{figure*}

To illustrate MPS algorithms for plasma dynamics we oultine the 1D2V bump on tail instability test case considered in Ref.~\cite{Plasma_Connor}. This test case was implemented with dynamical electric fields, and it was observed that the singular values of the encoded distribution plasma function decayed in a rapid exponential manner, suggesting strong compatibility with MPS approaches. While not directly quantified, this suggests that substantial compression could be achieved in this case, whilst avoiding the introduction of statistical noise. We show the dynamic snaphots of the electron distribution of this test case in Fig.~\ref{fig:Bump_snaps}, where we observed the formation of vortices in phase space as expected.

\begin{figure}[thb!]
    \centering
    \includegraphics[width=0.55\linewidth]{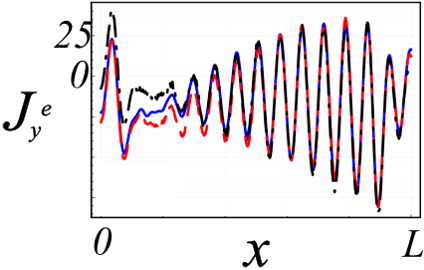}
    \caption{Demonstration of convergence of MPS simulations of the beat wave problem reproduced from \cite{Plasma_Connor}. We plot a cut along the x-y plane at $y=L_x/2$ of the current density from the electrons, $J_y^e(x,y)$ at a fixed simulation time of $t=0.14$. We show convergence of the MPS simulations for bond dimensions of $\chi=150$ (dash dotted black), $\chi=200$ (dashed red) and $\chi=512$ (solid blue).  }
    \label{fig:Jy_conv}
\end{figure}

Another example is the beat wave test case considered in Ref.~\cite{Plasma_Connor}. In this setup two opposing laser beams with a mismatch frequency are injected into a rectangular slab of uniform plasma and left to evolve. After the lasers overlap, a beating pattern develops in the electric field, which was validated against a PIC simulation. This system was first encoded into an MPS framework, and the impact of restricting bond dimension onto plasma observables was studied. It was found that substantial bond dimensions of $\chi=512$ was required to reproduce the current density without noticeable truncation errors, as shown in Fig.~\ref{fig:Jy_conv}. 

While it has been shown that changing the MPS topology to that of a comb tensor network \cite{ye2024quantized}, shown in Fig.~\ref{fig:Beat_encodings} can provide further compression over the MPS by $>4\times$ \cite{Plasma_Connor}, we will focus on the case of a standard MPS, with the method developed in Sec.~\ref{sec:new} being applicable for any tensor topology.

\section{Arbitrary injection of electromagnetic fields}
\label{sec:new}

\begin{figure}[t!]
     \centering
     \includegraphics[width=0.7\linewidth]{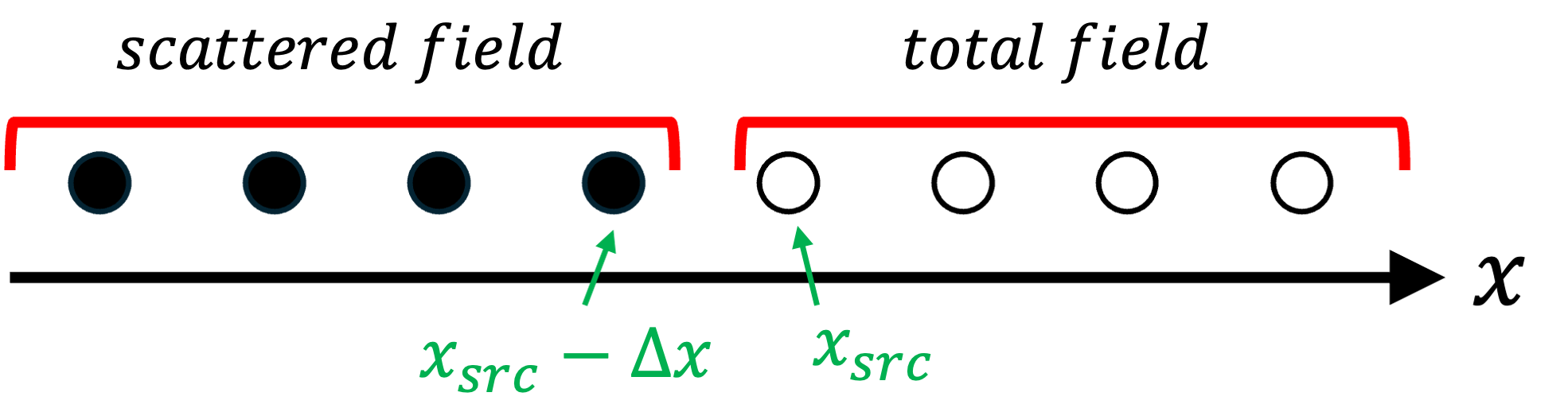}
     \caption{Pictorial representation of the total field / scattered field (TFSF) approach for electric fields. The simulation domain is broken into a total field region, where the electric field from the laser and all reflections and self-generated fields are included, and a scattered region without the injected laser field. The laser is defined to be injected along the boundary between the scattered and total regions. We show a sample discretisation along the $x$ axis, with grid-points shown as circles, and highlight the grid-points $x_\mathrm{src}-\Delta x$ and $x_\mathrm{src}$ which must be modified for calculation of derivatives.}
     \label{fig:TFSF}
\end{figure}

We now advance the MPS approach to enable arbitrary injection of electromagnetic fields, e.g., lasers of various profiles, into a plasma, as can be done with PIC methods. While previous work did consider the injection of electromagnetic fields, this was done in a constrained way such that fields could only be introduced at a boundary of the plasma \cite{Plasma_Connor}.

An extension to laser injection is the implementation of the total field scattered field (TFSF) approach. The core concept of TFSF is to split apart the simulation domain into a region in which the total electric field is specified, and a region in which only the electric field scattered from the bulk is known, illustrated in Fig.~\ref{fig:TFSF}a. In the bulk of both regions, Maxwells equations can be solved using standard finite difference approaches. At the boundary of the total and scattered region a laser is injected. In the scattered field region, there is no contribution from the laser as we assume it only propagates forward. Let us consider a central finite-difference scheme approximation for the spatial derivative for the point $x_\mathrm{src}-\Delta x$, i.e. the grid point before the source term. The grid-point at $x_\mathrm{src}-\Delta x$ resides in the scattered region, whilst $x_\mathrm{src}$ is in total field region. As we wish to calculate the scattered field, $E^S$, at $x_\mathrm{src}-\Delta x$, we must subtract from the total field, $E^T$, at $x_\mathrm{src}$ the analytical form of the injected laser, $E^L$. The derivative of the scattered field is
    \begin{equation}
        \hat{\partial}_xE^S|_{x_\mathrm{src}-\Delta x}=  \frac{E^{T}(x_\mathrm{src})- E^{L}(x_\mathrm{src})-E^{S}(x_\mathrm{src}-2\Delta x)} {2\Delta x}  \ ,
    \end{equation}
and hence one must add a contribution of $-\frac{E^{L}}{2\Delta x}$ to the derivative of the field at the grid point $x_\mathrm{src}-\Delta x$. Likewise for the first total field grid-point at $x_\mathrm{src}$ the derivative for the total field must be modified.  One can perform this modification via defining two tensor-networks $L$ and $R$ which are zero everywhere except at the grid-points of $x_\mathrm{src}-\Delta x$ and $x_\mathrm{src}$ respectively. The net derivative of the field is obtained by contraction of the chosen derivative MPO $\hat{\partial}_x$ with these MPS,
    \begin{equation}
        \frac{\partial E}{\partial x}= \hat{\partial}_x E - \frac{E^{L}}{2\Delta x}|_{x_\mathrm{src}} L - \frac{E^{L}}{2\Delta x}|_{x_\mathrm{src}-\Delta x} R \ .
    \end{equation}
Such an approach can be extended to a system of two scattered regions to allow for the injection of two opposing laser beams. 

\begin{figure}[t!]
     \centering
     \includegraphics[width=\linewidth]{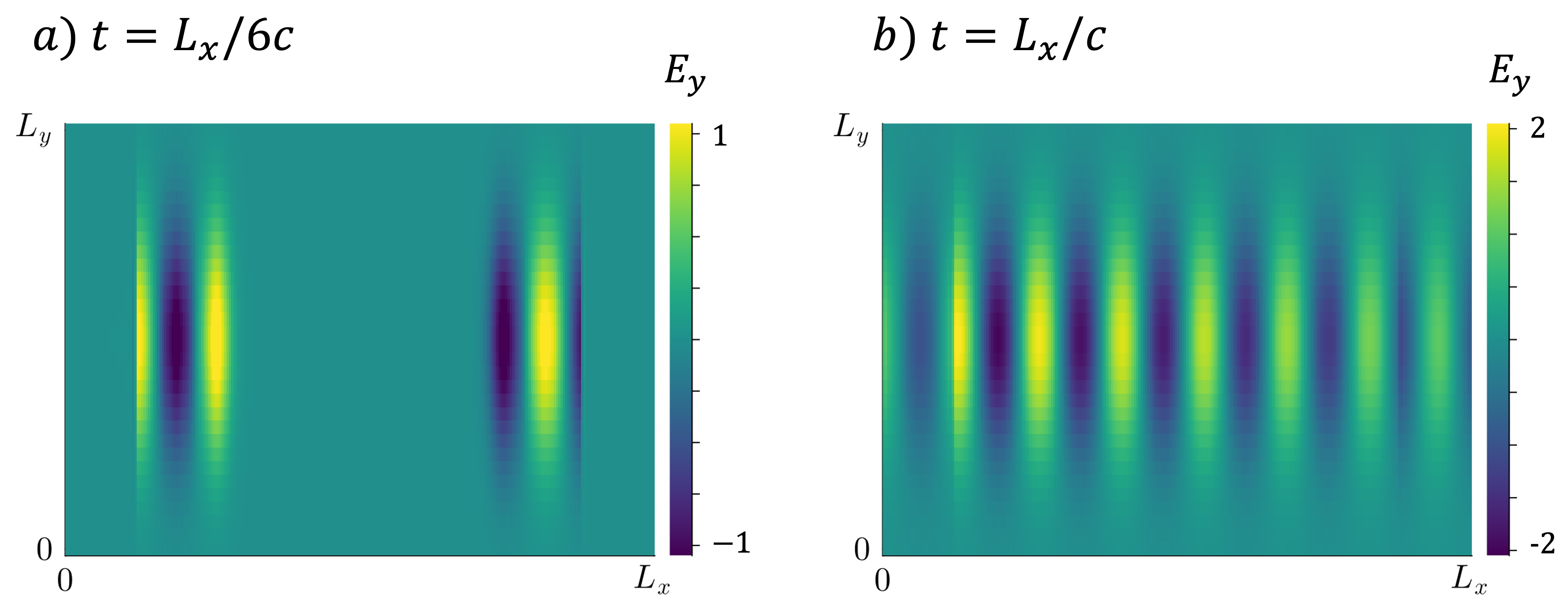}
     \caption{Snapshot of the $E_y$ electric field component from an injected laser field at times of $a)t=L_x/6c$ and $b) t=L_x/c$ making use of the TFSF approach. We choose the total field region as the central region, with the scattered field regions consisting of the $32$ grid-points nearest the boundaries at $x=0$ and $x=L_x$. The lasers by definition are injected at the boundaries between the total and scattered field regions. These simulations were conducted entirely within the MPS framework using the technique outlined in the text. }
     \label{fig:TFSF_Ey}
 \end{figure}

As an illustration of the combined MPS and TSTF approach, we simulate the propagation of two opposing lasers via Maxwells equations in the MPS formalism. We inject the laser profile at $32$ grid points into the grid, and calculate the propagation of the electromagnetic fields throughout the simulation region over time. The dynamics at two snapshots are shown in Fig.~\ref{fig:TFSF_Ey}, with the electric field in the $y$-direction plotted. We observe the correct dynamics of the electric field propagating from the injected point throughout the simulation region at later times.

\section{Outlook}
Tensor-network based methods remain a promising tool to perform large scale simulations of plasma dynamics in regimes where standard approaches are computationally infeasible or suffer from other limitations such as statistical noise and self-heating. The implementation of a TFSF approach for injected electromagnetic fields takes the tensor-network approach a step closer to being able to perform challenging industrial simulations with complex laser interactions as is currently handled by PIC methods. 

\begin{figure}[t!]
     \centering
     \includegraphics[width=.6\linewidth]{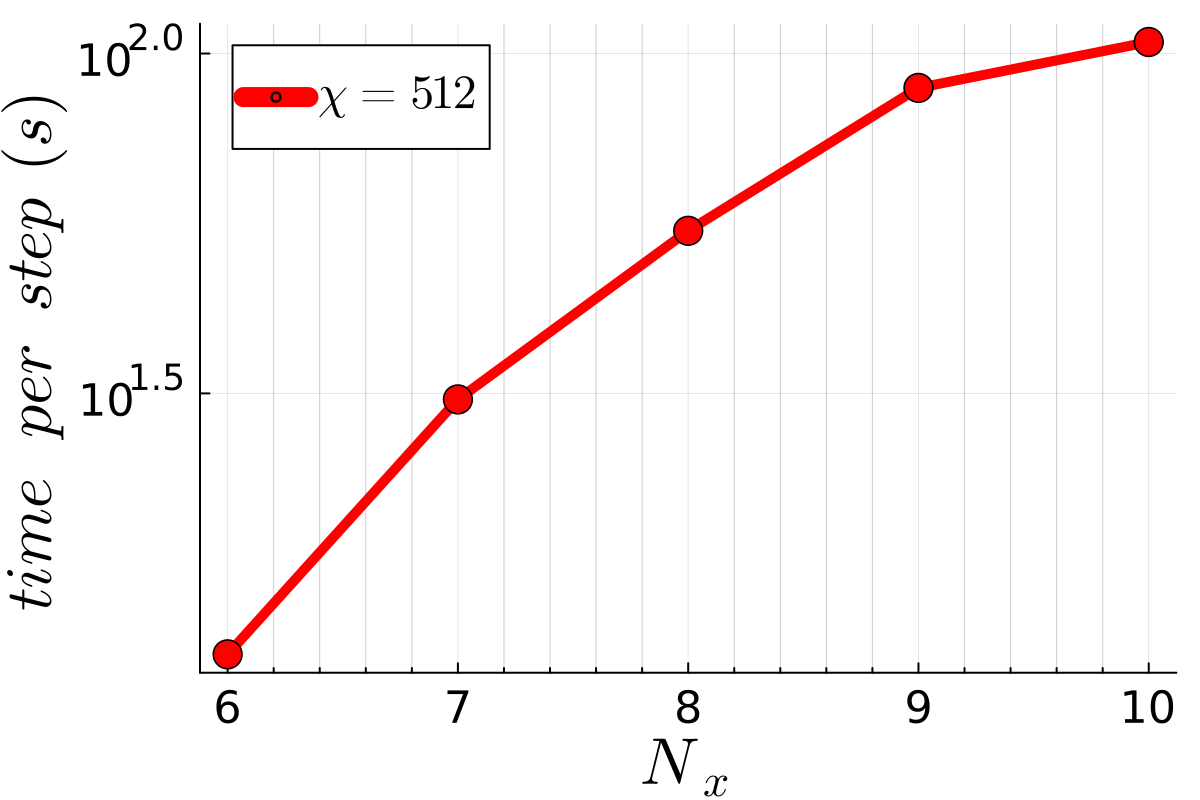}
     \caption{Demonstration of the computational scaling of tensor-network methods for plasma dynamics. We plot the average time required for the first $50$ time steps of the beat wave simulation of \cite{Plasma_Connor} with $\chi=512$ but making use of the TFSF approach. }
     \label{fig:scaling_plasma}
 \end{figure}

By making use of tensor-network approaches for plasma physics, we aim to take advantage of the improved computational resource scaling with respect to system size as compared against conventional approaches, whilst avoiding the need for statistical based methods. As a demonstration of this, we implemented the beat wave test case \cite{Plasma_Connor} using the TSTF approach in MPS representation, with the scaling of the runtime per step shown in Fig~\ref{fig:scaling_plasma}. The reuslts show that the introduction of capability to include arbitrary injection of electromagnetic fields has not destroyed the logarithmic scaling of the MPS approach with grid resolution. Note, the number of grid points is proportional to the logarithm of the number of tensors plotted on the x-axis of Fig.~\ref{fig:scaling_plasma}.

Further work is needed to increase the capabilities of the MPS approach to replicate the mature status of PIC methods, and other approaches to plasma dynamics and non-linear partial differential equations. However, the MPS approach is showing great promise through the variety of test cases now studied and the introduction of new capabilities, including the arbitrary electromagnetic field injection introdcued and demonstrated here.

UK Ministry of Defence \copyright Crown owned copyright ~\copyright2026 /AWE.



\end{document}